\documentclass[floats,aps,showpacs,twocolumn,groupedaddress]{revtex4}

\usepackage[dvips]{graphicx}
\usepackage{amsfonts}
\usepackage{amsmath,amssymb}
\usepackage{dsfont}
\usepackage{placeins}
\usepackage{afterpage}
\usepackage{flafter}     

\newcommand{\ud}{\text{d}}
\newcommand{\ui}{\text{i}}
\newcommand{\ue}{\text{e}}

\newcommand{\R}{\mathbb{R}}

\newcommand{\Z}{\mathbb{Z}}

\newcommand{\psireg}{\psi_{\text{reg}}}
\newcommand{\psiregmn}{\psi_{\text{reg}}^{mn}}
\newcommand{\psich}{\psi_{\text{ch}}}
\newcommand{\Hreg}{H_{\text{reg}}}
\newcommand{\neff}{n_{\text{eff}}}
\newcommand{\rhoch}{\rho_{\text{ch}}}
\newcommand{\Ach}{A_{\text{ch}}}

\newcommand{\Real}{\text{Re}}
\newcommand{\Imag}{\text{Im}}

\newcommand{\Qdyn}{Q_{\text{dyn}}}
\newcommand{\Qdir}{Q_{\text{dir}}}

\begin{document}

\title{Quality factors and dynamical tunneling in annular microcavities}

\author{Arnd B\"acker}
\author{Roland Ketzmerick}
\author{Steffen L\"ock}
\affiliation{Institut f\"ur Theoretische Physik, Technische Universit\"at
             Dresden, D-01062 Dresden, Germany}
\author{Jan Wiersig}
\affiliation{Institut f\"ur Theoretische Physik, Universit\"at
             Magdeburg, Postfach 4120, D-39016 Magdeburg, Germany}
\author{Martina Hentschel}
\affiliation{Max-Planck-Institut f\"ur Physik komplexer Systeme,
             N\"othnitzer Stra{\ss}e 38, D-01187 Dresden, Germany}

\date{\today}

\begin{abstract}
The key characteristic of an optical mode in a microcavity is its quality
factor describing the optical losses. The numerical computation of this
quantity can be very demanding for present-day devices. Here we show for a
certain class of whispering-gallery cavities that the quality factor is
related to dynamical tunneling, a phenomenon studied in the field of quantum
chaos. We extend a recently developed approach for determining
dynamical tunneling rates to open cavities. This allows
us to derive an analytical formula for the quality factor which is in very
good agreement with full solutions of Maxwell's equations.
\end{abstract}
\pacs{42.55.Sa, 42.60.Da, 05.45.Mt}

\maketitle

\noindent

\section{Introduction}
Optical microcavities in which photons can be confined in three
spatial dimensions are a subject of intensive research as they are relevant for
applications, such as ultralow-threshold lasers
\cite{ParKimKwoJuYanBaeKimLee2004,UlrGieAteWieReiHofLoeForJahMic2007},
single-photon emitters \cite{KimBenKanYam1999,MicImaMasCarStrBur2000}
or correlated photon-pair emitters \cite{BenUlrMicWieJahFor2005}.
Especially whispering-gallery cavities such as microdisks
\cite{MccLevSluPeaLog1992,MSJPLHKH07,KKPV06},
microspheres \cite{ColLefBruRaiHar1993,GMM06}, and microtoroids~\cite{AKSV03}
have been investigated as they can trap photons for a long time near the
boundary by total internal reflection. The corresponding whispering-gallery
modes have a very high quality factor $Q$, which makes these cavities a
candidate for the above-mentioned devices.
While the microdisk emits the photons isotropically, cavities with deformed
surfaces may additionally lead to directed
emission~\cite{LSMGPL93,ND97,GCNNSFSC98,KLRK04,KTMJCC04,WieHen2006,THFH07,WieHen2008,SCL08,WildePhD,YWD09,HenKwo2009}.
A particularly interesting geometry is the annular cavity~\cite{HN97,HR02,SWH04}
-- a microdisk with a circular-shaped inclusion. A non-concentric (air)
hole as inclusion 
allows for unidirectional emission and high quality factors simultaneously \cite{WieHen2006} which 
for most applications are desirable.

\begin{figure}[tb]
  \begin{center}
    \includegraphics[angle = 0, width = 85mm]{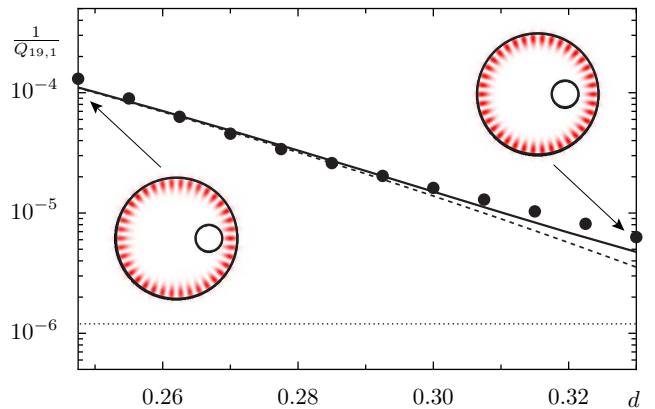}
    \caption{(Color online) Inverse quality factors $1/Q$ for the annular
             microcavity with refractive index $\neff=2.0$. Shown is the
             theoretical
             prediction (solid line), which is the sum of the direct
             tunneling contribution (dotted line)
             and the dynamical tunneling contribution
             (dashed line, Eq.~\eqref{eq:sternstern}),
             and numerical data
             (dots) for angular quantum number $m=19$ and
             radial quantum number $n=1$ vs the hole position $d$.
             The insets show the resonant state at $d=0.2475$ and $d=0.33$.
         }
    \label{fig:results_neff_2.0_var_d_19_1}
  \end{center}
\end{figure}

In this paper we connect the quality factors $Q$ of optical microcavities
to the concept of dynamical tunneling \cite{DavHel1981}.
We provide an explicit prediction for the quality
factors of whispering-gallery modes in the
annular microcavity; see Fig.~\ref{fig:results_neff_2.0_var_d_19_1}.
Microcavities typically have a mixed phase space where regions of regular
and chaotic motion coexist.
Dynamical tunneling occurs between these dynamically separated
phase-space regions.
While for one-dimensional systems
the tunneling process through an energetic barrier is well understood,
e.g., by means of WKB theory \cite{Mer1998}, dynamical tunneling
is a subject of intensive research experimentally
\cite{SteOskRai2001,HenHafBroHecHelMcKMilPhiRolRubUpc2001,FroWilHayEavSheMiuHen2002}
as well as theoretically \cite{HanOttAnt1984,Wil1986,BohTomUll1993,BBEM93,TomUll1994,ShuIke1995,
DorFriComb,HN97,Tom1998,Cre1998,PodNar2003,PodNar2005,SheFisGuaReb2006,BaeKetLoeSch2008,
BaeKetLoeRobVidHoeKuhSto2008}. 
Using a fictitious integrable system \cite{PodNar2003},
recently an approach has been developed 
\cite{BaeKetLoeSch2008,BaeKetLoeRobVidHoeKuhSto2008}
which successfully predicts dynamical tunneling rates for
quantum maps and billiards.
In this paper we extend this approach to open optical microcavities,
in particular to the annular cavity, to predict quality factors $Q$ of
whispering-gallery modes.

This paper is organized as follows. In Sec.~\ref{ch:cavities} we introduce
the annular cavity.
In Sec.~\ref{ch:quality-factors} we define the quality factors
and derive the dynamical-tunneling contribution, which is
then compared with numerical results.
A summary is given in Sec.~\ref{ch:summary}.

\section{Annular Microcavities}\label{ch:cavities}

Optical cavities are described by Maxwell's equations which in the case of
quasi-two-dimensional microdisks reduce to
a two-dimensional scalar mode equation~\cite{Jackson83eng}
\begin{equation}\label{eq:wave}
-\nabla^2\psi = n^2(x,y)k^2\psi
\end{equation}
with (effective) index of refraction $n(x,y)$, wave number $k=\omega/c$, frequency
$\omega$, and the speed of light in vacuum $c$.
The mode equation~(\ref{eq:wave}) is valid for both transverse magnetic (TM)
and transverse electric (TE) polarization. We focus on TM
polarization with the electric field $\vec{E}(x,y,t) =
(0,0,\psi(x,y)e^{-i\omega t})$
perpendicular to the cavity plane.
The wave function $\psi$ and its normal derivative are continuous across the
boundary of the cavity. At infinity, only outgoing-wave components are allowed.

The mode equation~(\ref{eq:wave}) with the above-mentioned boundary conditions
has analytical solutions only for special geometries such as the circle
(see the Appendix) or several concentric circles.
General geometries require numerical schemes such as the boundary element
method~\cite{Wiersig02b} which we use in this paper. In this approach
the two-dimensional partial
differential equation~(\ref{eq:wave}) is rewritten as a one-dimensional integral
equation involving Green's functions. Outgoing-wave conditions can be easily
fulfilled by using the outgoing solution for the Green's functions. The
boundary element method turns out to be very efficient even for computing
highly excited modes and their quality factors.

\begin{figure}[tb]
  \begin{center}
    \includegraphics[angle = 0, width = 85mm]{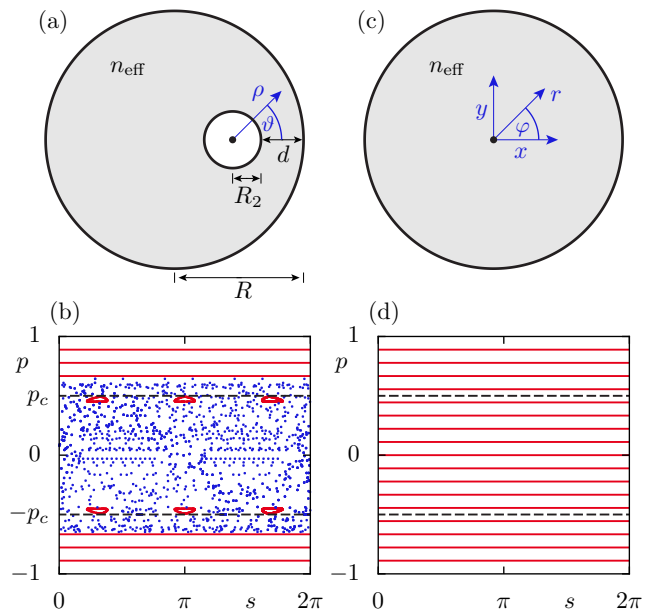}
    \caption{(Color online) (a) Annular cavity with $R=1$, $R_2=0.22$,
             and $d=0.33$. (b) Corresponding
             Poincar\'e section $(s,p=\sin\chi)$ for $\neff=2.0$, where
             $s$ is the arclength along the boundary and $\chi$ is
             the angle of reflection.
             Between the critical lines (dashed)
             with $|p_c|=1/\neff$
	           is the leaky region, where the condition
             for total internal reflection is not fulfilled.
             (c) Circular cavity and (d) its completely
             regular phase space.}
    \label{fig:cavity}
  \end{center}
\end{figure}

A particularly suited example to study the influence of a mixed phase space
onto the quality factors $Q$ is the annular cavity. Its geometry is given by
the radius $R$ of the large disk, the radius $R_2$ of the small disk and
the minimal distance between the two disks $d$, see Fig.~\ref{fig:cavity}(a).
Without loss of generality we choose $R=1$.
Under the variation in the
two parameters $d$ and $R_2$ the dynamics inside the cavity changes drastically
from completely regular behavior when the two disks are concentric ($d=R-R_2$)
to mixed regular-chaotic behavior for the general eccentric case.
This is clearly visible for different trajectories in the
Poincar\'e section, see Fig.~\ref{fig:cavity}(b).
The Poincar\'e section is a two-dimensional phase-space
representation. Whenever the trajectory hits the cavity's boundary,
its position $s$ (arclength coordinate along the circumference) and
tangential momentum $p = \sin{\chi}$ (the angle of reflection $\chi$ is
measured from the surface normal) is recorded.
The large disk has an effective refractive index $\neff$ while inside the
small disk and outside of the cavity the refractive index is unity.
For the visualization of the ray dynamics we
used an annular cavity with hard wall boundary conditions at the
outer disk, neglecting ray-splitting effects.

The annular cavity has been studied extensively in the context of quantum chaos~\cite{HR02},
optomechanics~\cite{SWH04}, avoided resonance crossings and resonant
tunneling~\cite{HN97}. For applications it is of high interest
as it allows for unidirectional light emission from high-$Q$
modes which has been predicted by the authors in Ref.~\cite{WieHen2006} and
has been confirmed in recent experiments~\cite{WildePhD}.
The closed system with perfectly reflecting walls, i.e., the annular billiard,
is a paradigm for dynamical tunneling~\cite{BBEM93,DorFriComb}.

\section{Quality factors} \label{ch:quality-factors}

The quality factor $Q$ of a mode in an open cavity is related to
the corresponding resonance with
complex wave number $k=\Real(k)+\ui\, \Imag(k)$ via
\begin{equation} \label{eq:Q-factor}
  Q=-\frac{\Real(k)}{2\Imag(k)} .
\end{equation}
In the annular cavity the quality factor $Q$ of a regular mode
has two contributions which are assumed to be additive
\begin{equation} \label{eq:epsilon}
  \frac{1}{Q}  = \frac{1}{\Qdir} + \frac{1}{\Qdyn}.
\end{equation}
Here, $\Qdir$ accounts for the direct coupling of the regular
mode to the continuum, as in the case of the circular cavity, 
see the Appendix. 
Note that for this contribution the mixed phase-space structure induced
by the small disk is irrelevant.
The second contribution, $\Qdyn$, is given by dynamical tunneling
from the regular mode to the chaotic sea, which for $|p|<p_c$ is
strongly coupled to the continuum, see Fig.~\ref{fig:cavity}(b).
Here we assume that there are no further phase-space structures within
the chaotic sea that affect the quality factor.
A priori it is not obvious, which of these contributions will dominate.

\subsection{Dynamical tunneling contribution}\label{ch:tunneling}

We now want to derive a prediction for the
dynamical tunneling contribution $\Qdyn$ of regular modes in
optical microcavities. After presenting the general approach
we will apply it to the annular microcavity.

\subsubsection{General approach}

We first review a quantum mechanical approach for determining dynamical
tunneling rates using a fictitious integrable system
\cite{BaeKetLoeSch2008,BaeKetLoeRobVidHoeKuhSto2008};
the relation to the quality factor $\Qdyn$ will be given below.
The tunneling rate $\gamma$ of a regular state to the chaotic 
sea is described by Fermi's golden rule (using units $\hbar=2M=1$)
\begin{equation}
 \label{eq:gamma_FGR}
 \gamma = 2\pi\langle|v|\rangle^2 \rhoch
\end{equation}
where $\rhoch$ denotes the density of chaotic states and $\langle|v|\rangle^2$
is the averaged squared matrix element between the considered regular state and
the chaotic states of similar energy. According to the Weyl formula for closed
two-dimensional billiards the density
of chaotic states is given by $\rhoch\approx \Ach/(4\pi)$, where $\Ach$
is the area of the billiard times the fraction of the chaotic phase-space
volume. 
The eigenmodes of a system with a mixed phase space are 
mainly regular or chaotic, i.e., concentrated
on a torus inside the regular region or spread out over the
chaotic component.
To calculate the coupling matrix
elements $v$ these so-called regular and chaotic eigenmodes cannot be
used as they have small, but still too large, admixtures of the other type
of modes compared to the tunneling rate.
Instead, we determine $v$ by introducing a fictitious integrable
system $\Hreg$ as it was first suggested for dynamical tunneling in Ref.~\cite{PodNar2003}. 
$\Hreg$ has to be chosen such that its classical dynamics resembles the regular
dynamics of the mixed system as closely as possible and extends it to
phase space regions where $H$ has a chaotic sea.
The eigenstates $\psireg$ of $\Hreg$ are localized in the regular
region of $H$ and decay into the chaotic sea of $H$.
With chaotic states $\psich$, which live in the
chaotic region of phase space, the coupling matrix element is given as
\cite{BaeKetLoeSch2008,BaeKetLoeRobVidHoeKuhSto2008,BaeKetLoe:InPreparation}
\begin{equation}
 \label{eq:coupling_matrix_element_general}
 v = \int_{\R^2} \psich^*(x,y)(H-\Hreg)\psireg(x,y) \,\ud x \ud y.
\end{equation}
Note that this equation is applicable for general systems 
but the determination of a sufficiently 
accurate $\Hreg$ is a difficult task. As previously mentioned we 
assume that there are no further phase-space structures within
the chaotic sea that affect the tunneling rates $\gamma$.
This approach was previously used to predict tunneling rates
for quantum maps \cite{BaeKetLoeSch2008}
and closed billiard systems \cite{BaeKetLoeRobVidHoeKuhSto2008}.

The described approach can be extended to open cavities in the
following way: (i) as a fictitious integrable system $\Hreg$ we choose
a cavity such that it resembles
the regular dynamics of $H$. The quantum system has resonance states $\psireg$.
(ii) As a model for the
chaotic resonances $\psich$ a random wave model will be used,
which in addition fulfills the relevant cavity boundary conditions.
(iii) The tunneling rate $\gamma$ determines the quality factor $\Qdyn$ by
\begin{equation} \label{eq:stern}
  \Qdyn  =  \frac{2 \Real(k)^2 \neff^2}{\gamma}
\end{equation}
where we used the quantum mechanical relation between energy and momentum
$E_0-\ui\frac{\gamma}{2} = p^2$, with $p=\neff k$ in a
refractive medium such that
$ \gamma = -2 \neff^{2} \Imag(k^2) =  -4 \neff^{2} \Real(k) \Imag(k)$.

\subsubsection{Application to the annular cavity}

To evaluate Eq.~\eqref{eq:coupling_matrix_element_general}
we have to find an appropriate regular system $\Hreg$.
For the annular cavity a natural choice is given by the circular cavity
as it correctly reproduces the regular
whispering-gallery motion of the annular cavity and extends it into the chaotic
region of phase space.
The whispering-gallery modes are labeled by the two quantum numbers $m$ and
$n$.
Rewriting the mode equation~(\ref{eq:wave}) as an eigenvalue equation $H\psi =
k^2\psi$ the Hamiltonian of the annular cavity can be introduced as
\begin{eqnarray}
 \label{eq:H}
 H = -\nabla^2 + [1-n(x,y)^2]k^2
\end{eqnarray}
where the refractive index $n(x,y)$ is $\neff$ inside and $1$
outside the cavity and in the disk of radius $R_2$.
As the regular system $\Hreg$ we choose the circular cavity
\begin{eqnarray}
 \label{eq:Hreg}
 \Hreg = -\nabla^2 + [1-n_{\text{reg}}(x,y)^2]k^2
\end{eqnarray}
where the refractive index $n_{\text{reg}}(x,y)$ is $\neff$ inside
and $1$ outside the circular cavity.
Thus $H$ and $\Hreg$ differ only inside the small disk of radius $R_2$ with
$H-\Hreg = (1-\neff^2) k^2$ and the integral in
Eq.~\eqref{eq:coupling_matrix_element_general} reduces to an integral over
the small disk.
For the regular states inside the circular cavity we choose,
as in the numerical studies, the even eigenmodes
\begin{equation}
 \label{eq:psireg}
 \psiregmn(r,\varphi) = N_{mn} J_m(\neff k_{mn} r) \cos(m\varphi)
\end{equation}
where $k_{mn}$ are the complex resonant wave numbers, according to the Appendix,
and $\psiregmn$ is
normalized to one with the numerically determined
normalization constant $N_{mn}$.

To model the chaotic modes $\psich$ within the small disk we employ a
random wave description \cite{Ber1977},
which has been extended to systems with a mixed phase space \cite{BaeSch2002}.
While this model accurately describes the random behavior in a medium with
constant refractive index, it cannot account for the change in refractive index
at the border of the small disk at $\rho=R_2$. We extend a boundary-adapted
random wave model \cite{Ber2002} to account for this boundary condition.
This is essential especially for low $\Real(k)$ as then all chaotic modes decay
inside the small disk, which cannot be reproduced by the usual random wave model.
Therefore we construct the chaotic states $\psich$ as a random superposition
of modes of a circular cavity of radius $R_2$ with refractive
index $1$ which is surrounded by a medium with refractive index $\neff$.
As follows from Eq.~\eqref{eq:circular_cavity_radial_solution}
these modes, at a fixed complex wave number $k_{mn}$ and
within the small disk, are
\begin{equation}
 \label{eq:psich_single_contributions}
 \psi_l(\rho,\vartheta) = A_{l} J_{l}(k_{mn}\rho)\cos(l\vartheta), \quad \rho\leq R_2 .
\end{equation}
The chaotic states $\psich$ are then constructed
by a random superposition of these modes
\begin{equation}
 \label{eq:psich_sum}
 \psich(\rho,\vartheta) = \frac{1}{\sqrt{\Ach}}\sum_{l=1}^{\infty} a_l \psi_l(\rho,\vartheta).
\end{equation}
Here the coefficients $a_l$ are
Gaussian random variables with mean zero and
$\langle a_l a_k \rangle = \delta_{l,k}$.
The random waves, constructed in such a way, fulfill the normalization condition
$\langle|\psich|^2\rangle = 1/\Ach$ required for the annular cavity.

Using the fictitious regular system and the random wave model
for the chaotic states we obtain an integral over the small disk
for the coupling matrix element
\begin{equation}
 \label{eq:coupling_matrix_element_eval}
 v_{mn} = \int_{0}^{R_2} \int_{0}^{\pi} \rho\,\ud\rho\ud\vartheta\, \psich^{*}(\rho,\vartheta)
                  (1-\neff^2) k_{mn}^{2} \psireg(\rho,\vartheta).
\end{equation}
For the tunneling rate this results in
\begin{equation}
 \label{eq:gamma_final}
 \gamma_{mn} = \frac{1}{2} N_{mn}^{2} (1-\neff^2)^2 |k_{mn}^{2}|^{2} \sum_{l=1}^{\infty} |I_{l}|^{2}
\end{equation}
where
\begin{equation}
 \label{eq:gamma_final_integral}
 I_{l} = \int_{0}^{R_2} \int_{0}^{\pi} \rho\,\ud\rho\ud\vartheta\,
               \psi_{l}^{*}(\rho,\vartheta) J_m(\neff k_{mn} r)
               \cos(m\varphi)
\end{equation}
with $r=r(\rho, \vartheta)$ geometrically related to
$\varphi=\varphi(\rho, \vartheta)$, see Figs.~\ref{fig:cavity}(a) and (c).
With Eq.~\eqref{eq:stern} we finally obtain the dynamical
tunneling contribution to the quality factor
\begin{equation} \label{eq:sternstern}
  \Qdyn^{mn} =  \frac{4 \neff^2}{N_{mn}^{2}  (1-\neff^2)^2 |k_{mn}^{2}| \sum_{l=1}^{\infty} |I_{l}|^{2} }
\end{equation}
of
whispering-gallery modes in the annular microcavity
for each quantum number $m$ and $n$.

\subsection{Results}

Now we compare our theoretical prediction for the quality factor,
Eq.~\eqref{eq:epsilon},
for the annular microcavity with numerical data,
obtained using the  boundary element method~\cite{Wiersig02b}.
Figure~\ref{fig:results_neff_2.0_var_d_19_1}
shows the inverse quality factors at fixed quantum numbers
$m=19$ and $n=1$ under variation in the distance $d$ between the small and the
large disk. The direct contribution $1/\Qdir$,
see the Appendix, is independent of the distance $d$.
It is dominated by the dynamical tunneling contribution $1/\Qdyn$,
Eq.~\eqref{eq:sternstern}, which decreases exponentially with $d$,
as expected from the increasing regular phase space region.
We find excellent agreement of the prediction and the numerical data.
\begin{figure}[tb]
  \begin{center}
    \includegraphics[angle = 0, width = 85mm]{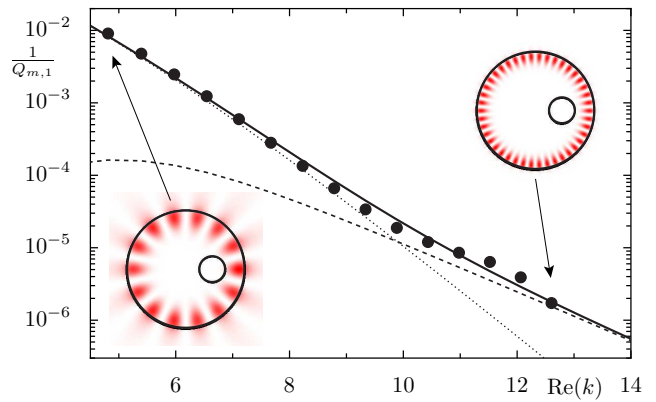}
    \caption{(Color online) Inverse quality factors $1/Q$ for the annular
             microcavity with $\neff=2.0$. Shown is the theoretical
             prediction (solid line) which is the sum of the direct
             tunneling contribution (dotted line) and the
             dynamical tunneling contribution
             (dashed line, Eq.~\eqref{eq:sternstern}), and numerical data (dots) 
             for $m = 7, \dots, 21$  and $n=1$ at $d=0.33$.
             The insets exemplarily show the resonant states of angular quantum number
             $m=7$ (left) and $m=21$ (right).
         }
    \label{fig:results_neff_2.0}
  \end{center}
\end{figure}

As a further test we consider the quality factors for fixed
radial quantum number $n=1$ and increasing angular quantum number $m=7, \dots, 21$,
comparing the theoretical prediction with numerical results,
see Fig.~\ref{fig:results_neff_2.0}.
We find that for small $\Real(k)$ the direct tunneling from the
whispering-gallery modes to the continuum is relevant while for large $\Real(k)$
the dynamical-tunneling contribution  dominates.
Here, our prediction again shows excellent agreement with the numerical data.
Note that for other systems, such as the one considered in Ref.~\cite{PodNar2005},
only $\Qdyn$ may be the relevant contribution.
Also we point out, that our theory allows to
determine quality factors for large $\Real(k)$, where numerical methods fail.
The boundary element method cannot compute
the quality factors of quantum numbers $n=1$ and $m > 21$ reliably as the
exponentially increasing quality factor requires an extremely fine spatial
discretization.

\subsection{Additional phase space structures}

In the derivation of the dynamical tunneling contribution $1/\Qdyn$
to the quality factor we assumed that there are no further structures
in the chaotic part of phase space, such as small regular islands and
partial barriers. If this assumption is not fulfilled,
the tunneling rate $\gamma$ is modified and consequently the quality factor.

\begin{figure}[tb]
  \begin{center}
    \includegraphics[angle = 0, width = 85mm]{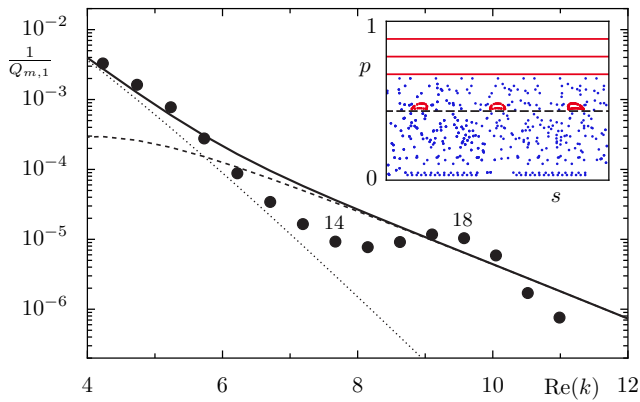}
    \caption{(Color online) Inverse quality factor $1/Q$ for the annular microcavity.
             Shown is the theoretical prediction (solid line)
             which is the sum of the direct
             tunneling contribution (dotted line) and
             the dynamical  tunneling contribution (dashed line),
             and numerical data (dots) for $m = 7, \dots, 21$, $n=1$
             at $d=0.33$ and $\neff=2.3$.
             The inset shows a Poincar\'e section of the classical phase space,
             where the critical line $p=p_c$ is marked (dashed line).
         }
    \label{fig:results_neff_2.3}
  \end{center}
\end{figure}

\begin{figure}[tb]
  \begin{center}
    \includegraphics[angle = 0, width = 85mm]{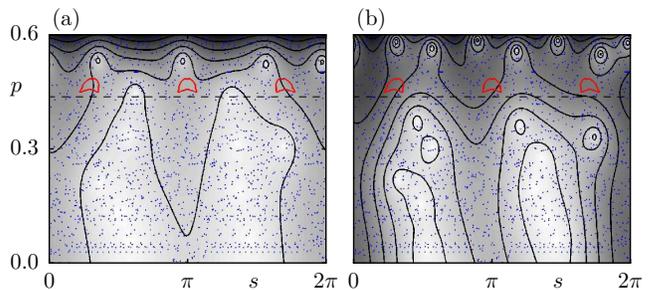}
    \caption{(Color online) Husimi functions (gray scale and contour lines) of
      the modes with angular quantum
      number (a) $m=14$ and (b) $m = 18$ superimposed onto part of the Poincar\'e
      section of the classical phase space,
      where the critical line $p=p_c$ is marked (dashed line).
      The radial quantum number is $n=1$ and
      the refractive index is $\neff=2.3$ as in Fig.~\ref{fig:results_neff_2.3}.
         }
    \label{fig:husimi}
  \end{center}
\end{figure}

This can be demonstrated when choosing $d=0.33$, $R_2=0.22$,
and increasing $\neff$ from $2.0$ to $2.3$. At $\neff=2.0$ no visible additional
structures exist in the chaotic part of phase space above the critical line
(see Fig.~\ref{fig:cavity}(b)).
At $\neff=2.3$
the critical line $p_c$ is shifted to smaller values and
a period-three island chain is now above $p_c$ as can be seen in the inset
in Fig.~\ref{fig:results_neff_2.3}. These structures presumably cause the
oscillations on top
of the numerically determined quality factors that are visible in
Fig.~\ref{fig:results_neff_2.3}. To support the conjecture, that the island
chain is responsible for the oscillations, Figs.~\ref{fig:husimi}(a) and
\ref{fig:husimi}(b) display the incident
Husimi functions~\cite{HSS03}, representing the wave analog of the
Poincar\'e section (see also \cite{BaeFueSch2004}),
of the mode $m=14$ (near the minimum of the oscillation) and
$m=18$ (near the maximum). In the former case the island chain is clearly a
barrier for the mode. The mode cannot penetrate the leaky region so
easily, which increases its quality factor. In the latter case the island
chain seems not to have a strong influence on the mode.
While the average behavior
of the quality factors is still well predicted by Eq.~\eqref{eq:sternstern},
these oscillations cannot be explained by our theory as it assumes a strong 
coupling of the chaotic modes to the continuum. 
Other situations where this coupling is weak are due to dynamical 
localization \cite{CasGuaShe1987} or to an additional tunnel barrier 
surrounding a cavity \cite{HN97}.
We leave the interesting
task of the prediction of tunneling rates through more complicated phase-space
structures in the chaotic sea as a future challenge.


\section{Summary}\label{ch:summary}
We have presented a theory for the intrinsic optical losses of annular
microcavities. It is assumed that the ray dynamical phase space is divided
into regular regions and a chaotic region which does not show additional
structures such as small regular islands or partial barriers. Our theory gives an
analytical
expression for the quality factor which is in very good agreement with the
full numerical simulations of Maxwell's equations. We would like to emphasize
that our theory can predict quality factors also in the regime of large wave
numbers where numerical methods fail due to the exponential
increase in the quality factor.

\begin{acknowledgments}
Financial support from the DFG research group 760 and the DFG Emmy
Noether Programme is acknowledged.
\end{acknowledgments}

\begin{appendix}

\section*{Appendix: The circular microcavity}\label{ch:circular_cavity}

For completeness, let us consider a circular microcavity of radius $a$ in more detail.
Inside the cavity the refractive index is denoted by $n_1$ and outside by $n_2<n_1$ (see Fig.~\ref{fig:cavity}(c)).
In the classical ray picture trajectories stay inside the cavity if their angle of incidence
with the boundary is larger than the angle of total internal reflection
$\arcsin(n_2/n_1)$. The dynamics is completely regular.
The circular cavity in TM-polarization is described by
the Schr\"odinger equation in polar coordinates $(r,\varphi)$
\begin{equation}
 \label{eq:SGL_circular_cavity}
 -\nabla^2\psi(r,\varphi) = n(r)^2 k^2 \psi(r,\varphi)
\end{equation}
where $n(r)$ changes from $n_1$ inside to $n_2$ outside the cavity.
The radial and the angular part can be separated, using the ansatz
$\psi(r,\varphi) = u(r)\phi(\varphi)$. We immediately obtain
$\phi(\varphi) = \ue^{\ui m\varphi}$, where $m\in\Z$ denotes the angular quantum
number. The radial part
\begin{equation}
 \label{eq:SGL_circular_cavity_radial}
 -\left(\frac{\partial^2}{\partial r^2}+\frac{1}{r}\frac{\partial}{\partial r}\right)
  u(r) + V_{\text{eff}}(r) u(r) = k^2 u(r)
\end{equation}
describes the motion of a particle in an effective potential
\begin{equation}
 \label{eq:circular_cavity_effective_potential}
 V_{\text{eff}}(r) = k^2[1-n(r)^2] + \frac{m^2}{r^2}.
\end{equation}
Metastable states inside this potential exist for $m/(n_1 a)<k<m/(n_2 a)$
and correspond to states with evanescent leakage that,
in the ray picture, are fully confined by total internal reflection.
The solutions of the radial equation are given as
\begin{equation}
 \label{eq:circular_cavity_radial_solution}
 u(r) = \left\{\begin{array}{ll}
                A_m J_m(n_1 k r), & r < a\\
                H_{m}^{(2)}(n_2 k r) + S_m H_{m}^{(1)}(n_2 k r), & r>a
               \end{array}\right.
\end{equation}
where the incoming wave is described by the Hankel function of the second kind
$H_{m}^{(2)}(n_2 k r)$ and the scattered one is described by the Hankel
function of the first kind $S_m H_{m}^{(1)}(n_2 k r)$ with a certain scattering
amplitude $S_m$. $A_m$ describes the amount of probability entering the cavity.
Using that the radial solutions and their derivative have to be continuous
at $r=a$ and that the scattering matrix has a pole at a resonance position,
this complex resonance position $k=\Real(k)+\ui\, \Imag(k)$ can be found by numerically
solving
\begin{equation}
 \label{eq:circular_cavity_resonance_equation}
 n_1 J_{m+1}(n_1 k a)H_{m}^{(1)}(n_2 k a) = n_2 J_{m}(n_1 k a)H_{m+1}^{(1)}(n_2 k a)
\end{equation}
for complex $k$. From this $\Qdir = - \Real(k)/[2 \Imag(k)]$ is obtained.

\end{appendix}

\end{document}